 \documentclass[]{aa}
\usepackage{graphics,epsfig,txfonts}
\usepackage{natbib}
\bibpunct{(}{)}{;}{a}{}{,}

\usepackage{graphicx}

\usepackage{longtable}
\usepackage{amssymb}
\usepackage{xcolor}
\usepackage{lipsum}
\usepackage{textcomp}
\usepackage[colorlinks=true,linkcolor=blue,citecolor=blue,
bookmarks=true,bookmarksopen=true,bookmarksnumbered=true,draft=false]{hyperref}
\usepackage{enumerate}
\usepackage{xspace}

\newcommand{\ergcm}[1]{$\times 10^{#1}$ erg cm$^{-2}$ s$^{-1}$}

\newcommand{\ergs}[1]{$\times 10^{#1}$ erg s$^{-1}$}
\newcommand{\oergs}[1]{$10^{#1}$ erg s$^{-1}$}
\newcommand{\hcm}[1]{$\times 10^{#1}$ cm$^{-2}$}

\newcommand{\expo}[1]{$\times 10^{#1}$}
\newcommand{\oexpo}[1]{$10^{#1}$}
\newcommand{\kms}{km s$^{-1}$}
\newcommand{\nh}{N$_{\rm H}$}

\newcommand{\ct}{cts s$^{-1}$}

\newcommand{\Halpha}{H${\alpha}$}
\newcommand{\ltsima}{$\buildrel < \over \sim$}
\newcommand{\lsim}{\lower.5ex\hbox{\ltsima}}
\newcommand{\gtsima}{$\buildrel > \over \sim$}
\newcommand{\gsim}{\lower.5ex\hbox{\gtsima}}

\newcommand{\rxp}{RX\,J0520.5-6932\xspace}
\newcommand{\swift}{{\it Swift}\xspace}

\newcommand{\xmm}{{\it XMM-Newton}\xspace}

\begin{document}
 
\title{Spectral and temporal properties of \rxp (LXP 8.04) during a type-I outburst.
\thanks{Based on observations with \xmm, an ESA Science Mission with instruments and contributions directly funded by ESA Member states and the USA (NASA); with Swift, a NASA mission with international participation.}
}

\author{G.~Vasilopoulos\inst{1} \and F.~Haberl\inst{1} \and R.~Sturm\inst{1}  \and  P.~Maggi\inst{1} \and A.~Udalski\inst{2}}

\titlerunning{Spectral and temporal properties of \rxp (LXP 8.04) during a type I outburst.}
\authorrunning{Vasilopoulos et al.}
 
\institute{Max-Planck-Institut f\"ur extraterrestrische Physik,
           Giessenbachstra{\ss}e, 85748 Garching, Germany\\
	   \email{gevas@mpe.mpg.de}
	   \and
           Warsaw University Observatory,
           Al. Ujazdowskie 4,
           00-478 Warszawa, Poland
	   }
 
  \date{Accepted 26/05/2014}
 
\abstract{}
{We observed \rxp in the X-rays and studied the optical light curve of its counterpart to verify it as a Be/X-ray binary.}
{We performed an \xmm anticipated target of opportunity observation in January 2013 during an X-ray outburst of the source in order to search for pulsations and derive its spectral properties. We monitored the source with \swift to follow the evolution of the outburst and to look for further outbursts to verify the regular pattern seen in the optical light curve with a period of $\sim$24.4 d.}
{The \xmm EPIC light curves show coherent X-ray pulsations with a period of 8.035331(15) s ($1\sigma$). The X-ray spectrum can be modelled by an absorbed power law with photon index of $\sim$0.8, an additional black-body component with temperature of $\sim$0.25 keV and an Fe K line. Phase-resolved X-ray spectroscopy reveals that the spectrum varies with pulse phase. We confirm the identification of the optical counterpart within the error circle of \xmm at an angular distance of $\sim$0.8 \arcsec, which is an O9Ve star with known \Halpha~ emission. By analyzing the combined data from three OGLE phases we derived an optical period of 24.43 d.}
{The X-ray pulsations and long-term variability, as well as the properties of the optical counterpart, confirm that \object{RX\,J0520.5-6932} is a Be/X-ray binary pulsar in the Large Magellanic Cloud. Based on the X-ray monitoring of the source we conclude that the event in January 2013 was a moderately bright type-I X-ray outburst, with a peak luminosity of 1.79\ergs{36}.}


\maketitle

\section{Introduction}
\label{sec-intro}

Be/X-ray binaries (BeXRBs) are a major subclass of high-mass X-ray binaries \citep[HMXBs, for a review see][]{2011Ap&SS.332....1R}. These binaries consist of a compact object, typically a neutron star (NS), that accretes matter from a Be star. Be stars are rapidly rotating main-sequence, sub-giant or giant stars typically having spectral classes between O5 and B9 \citep{2005MNRAS.356..502C}. These early-type stars are surrounded by a circumstellar (decretion) disk formed by viscous diffusion of matter ejected from the star. The disk can be truncated by the tidal/resonant interaction with the NS (e.g., \citealt{2001A&A...377..161O}). BeXRBs often have wide and eccentric orbits \citep{2011MNRAS.416.1556T}, although a population of low-eccentricity systems also exists and might indicate a distinct formation channel \citep{2002ApJ...574..364P}.

Generally in BeXRB systems, mass accretion strongly varies with orbital phase. Because of its eccentric orbit, a probable result of the supernova kick during the NS formation, or/and the possible misalignment between the inclination of the NS orbit and the orientation of the decretion disk, the NS can capture gas from the Be disk only for a short span of time during the close encounter with the Be disk. This brief interaction makes BeXRBs transient X-ray systems, mainly showing periodic (type I) outbursts ($L_{\rm x}\sim$\oergs{36-37}) correlated with their orbital period. However there are known persistent systems as well ($L_{\rm x}\lesssim$\oergs{34}). These are believed to be low-eccentricity systems where the mass accretion is relatively constant.
Sporadically, all BeXRBs might exhibit giant (type II) outbursts ($L_x\ge$\oergs{37}) that could be associated with warping episodes of the Be disk \citep{2013PASJ...65...41O}.
For some cases a temporary accretion disk can be formed around the NS, as indicated by the strong spin-up seen during bright outbursts.

BeXRBs may also show modulation of their X-ray flux with periods in the range of $\sim$1$-$1000 seconds. These X-ray pulsations are due to the misalignment of the NS’s rotation axis and its magnetic axis and provide one of the strongest evidence that the compact object is an NS. The strong magnetic field of the NS funnels mass accretion onto the magnetic poles and leads to the formation of an accretion column. Within these columns, shocks heat the gas, which then emits X-rays (for a detailed review of the emission mechanisms see \citealt{2004ApJ...614..881H,2002apa..book.....F}). The pulse profiles can be single peaked, double peaked or even more complex depending on the accretion geometry and viewing angle \citep{2010A&A...520A..76A}.

Temporal variation in brightness can be also observed from the optical counterpart of BeXRBs. It is the equatorial disk of circumstellar material around the Be star, which determines its observational properties (emission lines, infrared excess, variability). The long-term optical variability of the system can be attributed to the orbital period, disruption of the disk while the NS passes through the periastron, or other disk instabilities, like truncation or precession of the disk. In addition, some Be stars show intrinsic, short-period variations of low amplitude that are probably associated with non-radial pulsations (NRP) of the primary star, with periods between 0.2 and 1.7 days \citep{2008AJ....135.1350S}. \citet{2013MNRAS.431..252S} studied the 
optical light curves of BeXRBs in the Small Magellanic Cloud (SMC). They showed that 44\% of the systems exhibit signatures consistent with orbital periods, while 58 \% of them exhibit NRP.

The study of BeXRBs in nearby galaxies can provide more complete and homogeneous samples of BeXRB populations with the advantage of low foreground absorption and well known distances compared to sources in our Galaxy. The Magellanic Clouds (MCs) offer a unique possibility to study the population of high-energy sources of a complete galaxy.
With a moderate and well known distance of 50 kpc for the Large Magellanic Cloud (LMC) and 60 kpc for the SMC, 
faint sources down to a few $10^{33}$ erg s$^{-1}$ can be observed with \xmm\ in relatively short exposures.
Due to recent episodes of star formation, $\sim$40 Myr ago for the SMC and $\sim$(15$-$50)\,Myr for the LMC \citep{2010ApJ...716L.140A,2011AAS...21822829A}, the MCs harbour a large population of HMXBs.
Approximately 60 confirmed pulsars have been found in the SMC so far \citep{2004A&A...414..667H,2005MNRAS.356..502C,2010ASPC..422..224C}, and 45 candidates have been identified during the recent \xmm survey of the SMC \citep[][]{2013A&A...558A...3S,2012A&A...545A.128H}. On the other hand the LMC, which is 10 times more massive than the SMC, has only about 20 confirmed HMXB \citep[see][and references within]{2005A&A...442.1135L,2013A&A...554A...1M,2013A&A...558A..74V}, of which 14 are HMXB pulsars.
Still, since the X-ray coverage of the LMC is not nearly as complete as that of the SMC, early interpretations on the connection between star-formation rate and its BeXRB population would be dubious. Additionally, their transient behaviour complicates the investigation of BeXRBs, thus new systems are often discovered by chance. For constructing a more complete sample of their population, it is crucial to continuously monitor candidate systems or regions with an increased probability to find such systems (e.g. where recent star formation occurred) and perform deeper observations during outbursts \citep[e.g.][]{2013A&A...558A..74V,2013A&A...554A...1M,2012A&A...542A.109S}. 

In this work we report on a newly confirmed BeXRB in the LMC, \rxp \citep{2013ATel.4748....1V}. The source was discovered in X-rays by \citet{1994PASP..106..843S} with the ROSAT HRI. Subsequent ROSAT PSPC observations showed the source to be variable \citep{1999A&A...344..521H}. \citet{1994PASP..106..843S} proposed an O8e star found within the ROSAT 90\% error circle as the optical counterpart. \citet{2001MNRAS.324..623C} performed optical spectroscopy of the star and determined a spectral type of O9Ve (with one spectral subtype uncertainty). In the same paper, an analysis of the Optical Gravitational Lensing 
Experiment (OGLE) data revealed a 24.45 d optical periodic modulation of about 0.03 mag. \citet{2004MNRAS.349.1361E} confirmed this proposed optical period, based on the analysis of archival data from the Massive Compact Halo Objects (MACHO) project, while also reporting on the presence of an outburst lasting 200 d ($\sim$0.1 mag) around March 1995. From the same work, spectroscopic observations of the optical counterpart revealed the presence of a variable H$\alpha$ emission line, with an equivalent width of 5.2$\pm$0.2~\AA{} in 2001 November and 2.0$\pm$0.6~\AA{} in 2002 December.

The source was later detected by \xmm (2XMM J052029.7-693155) in an observation performed on 2004 January 17 (obsid: 0204770101), and in a recent observation performed on 2012 December 5 (obsid: 0690750901), as part of the \xmm survey of the LMC (PI: F. Haberl). From the above it is clear that while many of the optical properties of the system have been reported, its X-ray behaviour is still poorly studied and \rxp was mainly described as a weak X-ray 
source, until 2013 January when the source was detected by \swift, during the \swift UV survey of the LMC (PI: S. Immler),  while undergoing a moderately bright X-ray outburst \citep{2013ATel.4748....1V}. 

Following the first \swift detection, we further monitored the source and triggered an anticipated \xmm target of opportunity (ToO) observation (obsid: 0701990101, PI: R. Sturm) during the observed X-ray maximum flux. In the following paragraphs, we discuss the results of the spectral and timing analysis of the \xmm ToO observation, that resulted in the detection of coherent pulsations and a refined X-ray position. In Section \ref{sec-observations}, we describe  the X-ray data reduction and present the results of the spectral and temporal X-ray analysis. From the \swift monitoring we had indications that the system experienced several minor X-ray outbursts (possibly type I) that seem to repeat with the optical period of the system. In Section \ref{sec:datareduction_optical}, we present the updated optical light curve of the counterpart of \rxp.
A timing analysis using the combined optical light curve from phase II, III, and IV  of the OGLE survey resulted in a refined optical period. In Section \ref{discussion}, we discuss the results of the X-ray spectroscopy and summarise the main points that resulted from our analysis. 


\section{X-ray observations and data reduction}
\label{sec-observations}

\subsection{X-ray observations}   
\label{sec-xobs}
A \swift observation performed on 2013 January 13 as part of the LMC UV survey detected the HMXB candidate \rxp in a moderately bright X-ray outburst at a flux of 1.6$\pm$1.5\ergcm{-12} (0.3-10 keV). This is significantly brighter compared to the \xmm detection on 2012 December 5, which was at a flux level of 9.5$\pm$1.5\ergcm{-14} (EPIC-pn: 0.3-10 keV). Following the \swift detection we started to monitor \rxp with \swift to study the ongoing event that revealed a high X-ray variability of the source. After the proceeding rise of the source luminosity we triggered an \xmm ToO observation that was performed on 2013 January 22. We continued the \swift monitoring after the decline of the outburst, to study its long-term X-ray variability, and to search for a possible correlation between the X-ray and optical light curve. In total we acquired 22 \swift/XRT pointings, with a typical exposure of 2000 s (flux limit at $\sim$5\ergcm{-13}) within 130 
days. Apart from the data collected from our monitoring, additional six \swift/XRT fields from other surveys, that covered the position of \rxp, were used in our analysis. The complete log of the \swift and \xmm observations obtained and analysed for the current work is summarised in Table \ref{tab:xray-obs}.  

Regarding the \swift/XRT data reduction, we downloaded the data from the {\tt Swift Data Center}\footnote{http://swift.gsfc.nasa.gov/sdc/}. We estimated the source count rates by using the {\tt HEASoft}\footnote{http://heasarc.nasa.gov/lheasoft/} task {\tt ximage}, correcting for the vignetting and background effects, while we used the {\tt uplimit} task to estimate upper limits for the non-detections. For the brightest detections ($>$100 counts) we created X-ray spectra.  For the spectral extraction we used the {\tt HEASoft} task {\tt xselect} and circular regions with radius 35\arcsec~ for the source and 300\arcsec~ for the background.

In the \xmm ToO observation, the source was located on CCD4 of the EPIC-pn \citep{2001A&A...365L..18S} and on CCD1 of both the EPIC-MOS \citep{2001A&A...365L..27T} detectors. \xmm SAS 13.0.1\footnote{Science Analysis Software (SAS), http://xmm.esac.esa.int/sas/} was used for data processing. 
The background stayed at a low level for almost all the observing time and we excluded less than 2 ks of exposure due to a moderate flare seen in the EPIC-pn detector. 
We used a background threshold of 8 and 2.5 counts ks$^{-1}$ arcmin$^{-2}$ for the EPIC-pn and EPIC-MOS detectors, respectively. This resulted in net exposure times of 16.8/20.3/20.3 ks for EPIC-pn/MOS1/MOS2 after flare removal. The event extraction was performed using a circle around the source while the background events were extracted from a point-source free area on the same CCD but on different pixel columns for EPIC-pn. We optimised the size of the source extraction area using the SAS task {\tt eregionanalyse}. For the EPIC-pn spectra and light curves, we selected single-pixel and double-pixel events ({\tt PATTERN$\le$4}) while in the case of EPIC-MOS we used single to quadruple events ({\tt PATTERN$\le$12}). The quality flag {\tt FLAG = 0} was used throughout. The SAS task {\tt especget} was used to create the spectra and response files for the spectral analysis. The spectra were binned to achieve a minimum signal-to-noise ratio of five for each bin.

\subsection{X-ray position}
The position of \rxp was determined from the \xmm ToO observation that provided the best statistics. X-ray images were created from all the EPIC cameras using the \xmm standard energy sub-bands \citep{2009A&A...493..339W}. Source detection was performed simultaneously on all the images using the SAS task {\tt edetect\_chain}. Boresight correction was performed on the images based on the position of five identified background X-ray sources in the field. The positional correction based on these sources was found to be $\sim$1\arcsec. The final source position was determined to R.A. = 05$^{\rm h}$20$^{\rm m}$29\fs99 and Dec. = --69\degr31\arcmin55\farcs3 (J2000), with a $1\sigma$ statistical uncertainty of 0.05\arcsec. The total 1 $\sigma$ positional error, however is determined by the remaining systematic uncertainty assumed to be 0.5\arcsec \citep[see section 4.3 of][]{2013A&A...558A...3S}.

\subsection{Timing analysis}
\label{sec-time_an}

\begin{figure}
 \resizebox{\hsize}{!}{\includegraphics[angle=-90,clip=]{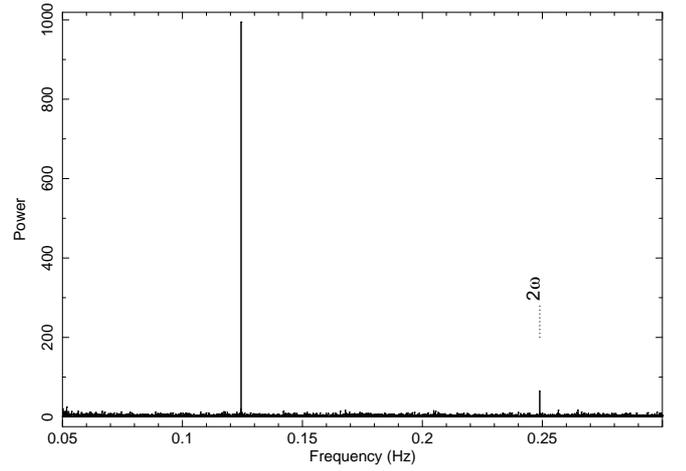}}
  \caption{Power density spectrum created from the merged EPIC-pn and EPIC-MOS data in the 0.2-10.0 keV energy band. 
           The time binning of the input light curve was 0.01 s. The best-fit frequency  $\omega\sim0.12445$ Hz and its first harmonic are easily distinguished in the plot.}
  \label{fig:psd}
\end{figure}

We used the SAS task {\tt barycen} to correct the \xmm EPIC event arrival times to the solar-system barycentre. To increase the signal-to-noise ratio we created time series from the merged event list of EPIC-pn and EPIC-MOS for the common good-time intervals. We used the {\tt HEASoft} task {\tt powspec} to search the X-ray light curve (0.2 -- 10 keV) for periodicities. In Fig.~\ref{fig:psd}, we present the inferred power density spectrum with a strong peak at the main period and its first harmonic. To improve the result of our initial estimation, we followed \citet{2008A&A...489..327H}. By using a Bayesian periodic signal detection method \citep{1996ApJ...473.1059G} we determined the pulse period with a 1$\sigma$ uncertainty to 8.035331(15) s. Following the nomenclature introduced by \citet{2005MNRAS.356..502C} for the BeXRB pulsars in the SMC we suggest the alternative name of \object{LXP\,8.04} for the system in the LMC.

\begin{figure}
  \resizebox{\hsize}{!}{\includegraphics[angle=0,clip=]{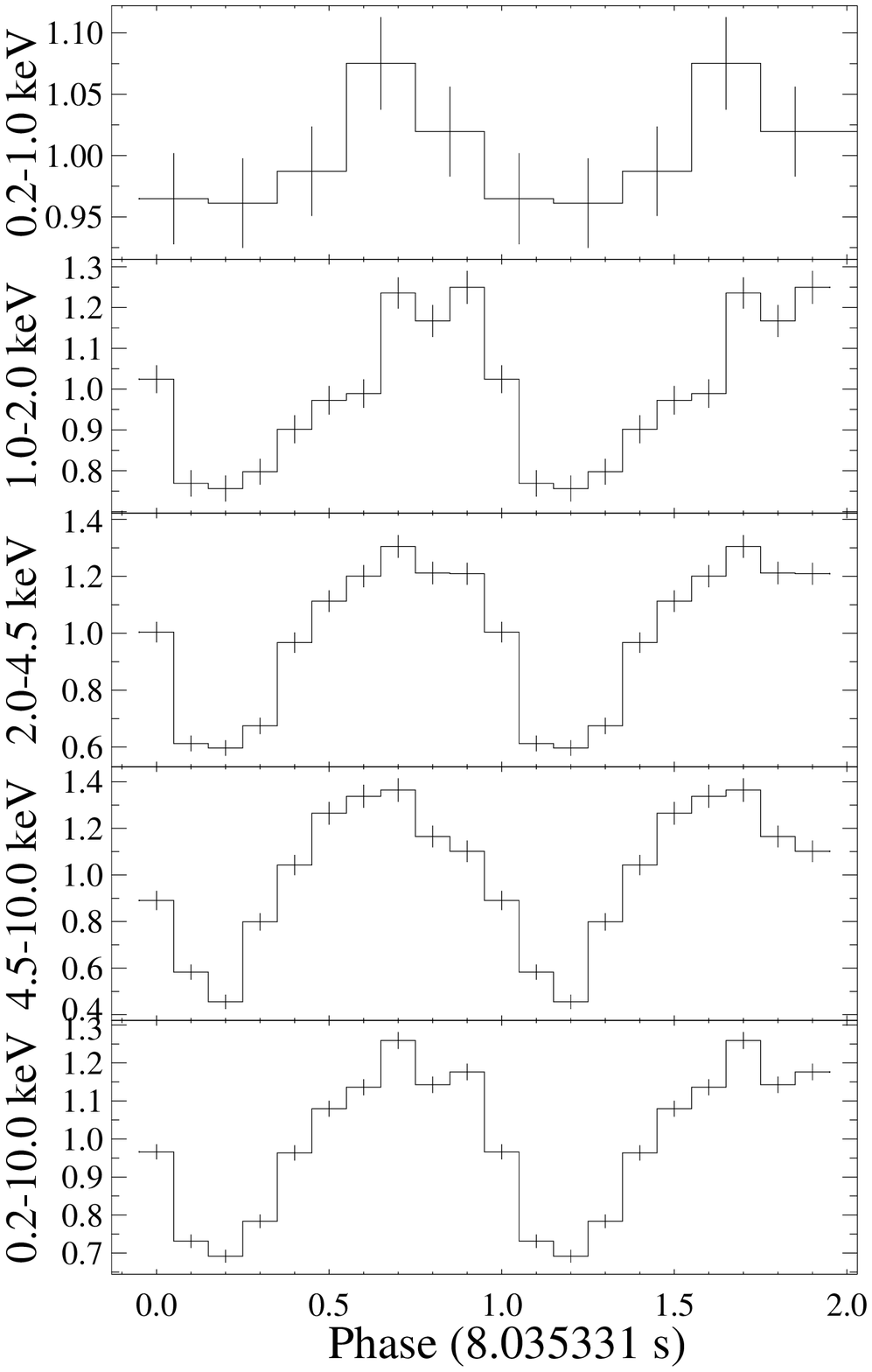}\includegraphics[angle=0,clip=]{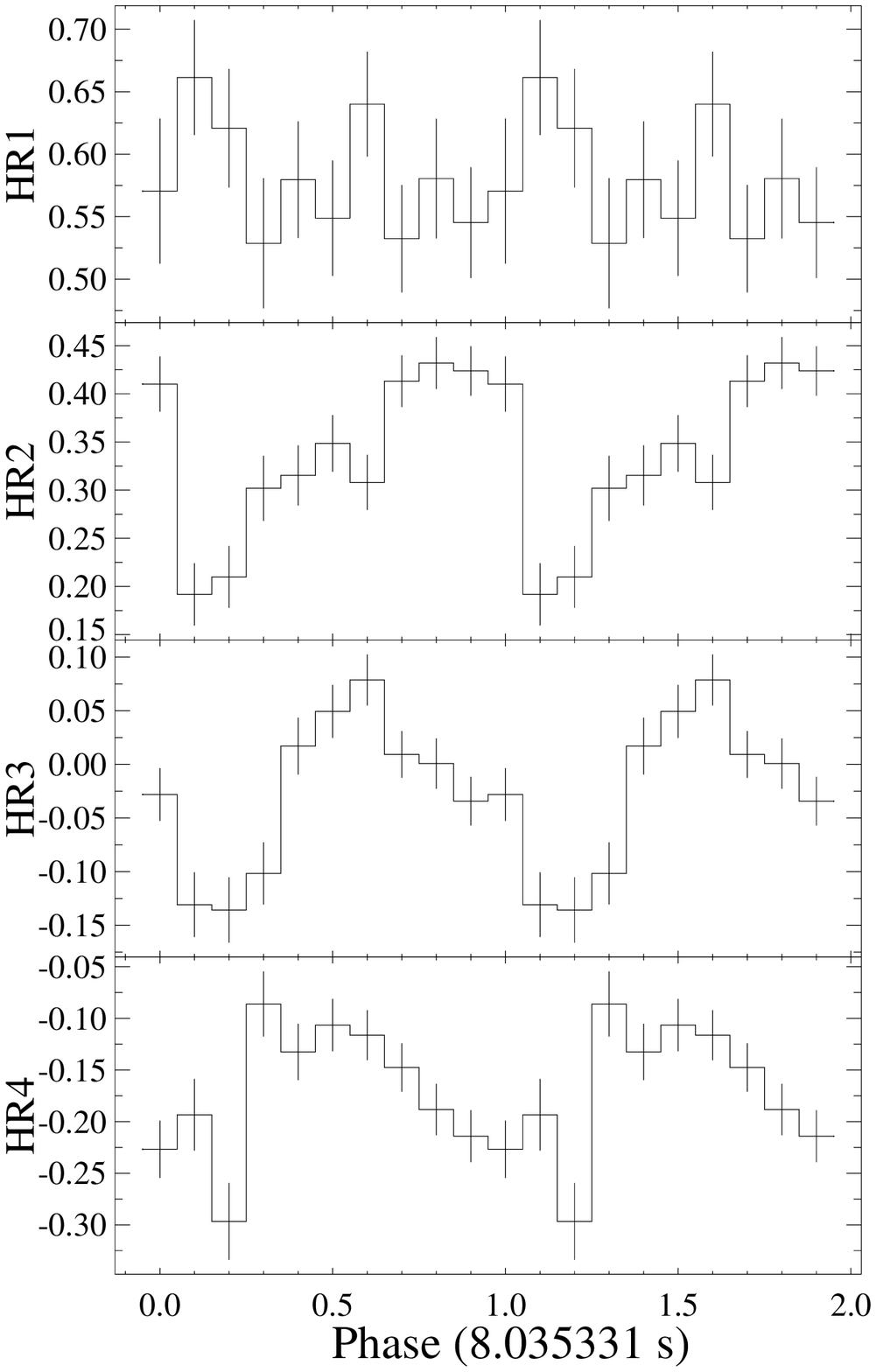}}
  \caption{Left: Pulse profiles obtained from the merged EPIC data in different energy bands (P=8.035331 s). All the profiles are background-subtracted and normalized to the average count rate (0.276, 0.454, 0.438, 0.311, 1.48 \ct, from top to bottom). 
Right: Hardness ratios as a function of pulse phase derived from the pulse profiles in two neighboring standard energy bands. 
	  }
  \label{fig:pp}
\end{figure}

By using the 5 standard energy bands (0.2-0.5, 0.5-1.0, 1.0-2.0, 2.0-4.5, 4.5-10 keV) we can define 4 hardness ratios as
$\rm{HR}_i=(\rm{R}_{\rm{i+1}}-\rm{R}_{\rm{i}})/(\rm{R}_{\rm{i+1}}+\rm{R}_{\rm{i}})$, with R$_{\rm i}$ denoting the background-subtracted count rate in energy band i. The period-folded pulse profiles in the EPIC standard energy bands together with the hardness ratios derived from the pulse profiles in two adjacent energy bands are plotted in Fig. \ref{fig:pp}. To achieve better statistics, the first two energy bands were combined in the top left panel, the bottom panel shows all five energy bands combined. 

From the HR diagrams, it seems that the pulse behaves different in the soft and hard X-ray bands. The source becomes first harder (see HR3, HR4 at phase $\sim$0.3$-$0.6) in the hard X-ray bands and after the pulse maximum, it becomes harder in the soft X-rays (see HR2). To further investigate this behaviour we created X-ray spectra for different phase intervals.

\subsection{Spectral analysis}
\label{sec-spec_x}
 \begin{figure}
    \resizebox{\hsize}{!}{\includegraphics[angle=-90,clip,trim=0 20px 0 0]{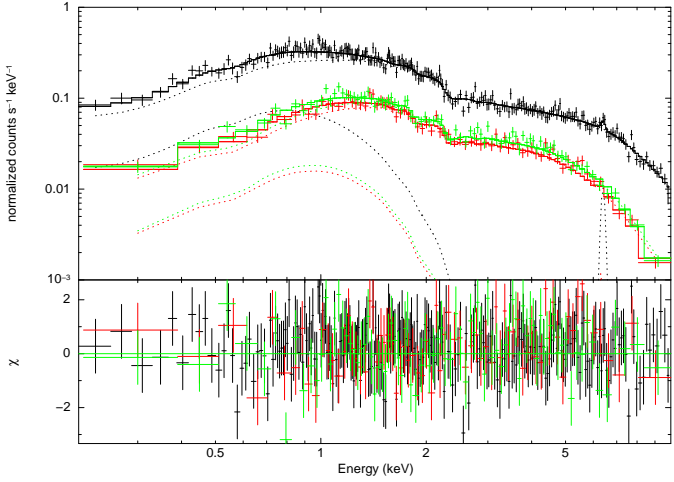}}
            \caption{Pulse-phase averaged EPIC spectra of \rxp. The top panel shows the EPIC-pn (black), EPIC-MOS1 (red) and EPIC-MOS2 (green) spectra, together with the best-fit model (solid lines) composed by an absorbed power law with photon index of 0.83, a black body with kT=0.24 keV (dotted lines) and a Gaussian line at 6.4 keV.
 	   The residuals for this model are plotted on the bottom panel.}
   \label{fig:spec1}
 \end{figure}

The spectral analysis was performed with {\tt xspec} \citep{1996ASPC..101...17A} version 12.8.0. The high number of counts enabled us to use $\chi^2$ statistics in the fitting procedure. The \xmm/EPIC spectra were fitted simultaneously using the same model parameters with an additional scaling factor to account for instrumental differences. For the EPIC-pn we fixed the scaling factor at 1 while for both EPIC-MOS we obtained values of $C_{MOS1}=1.03\pm0.03$ and $C_{MOS2}=1.00\pm0.03$, which is consistent with the expected values, as EPIC-MOS is known to provide $\sim$5\% higher fluxes than EPIC-pn
\citep[see ][or the latest version of the \xmm calibration manual\footnote{http://xmm2.esac.esa.int/external/xmm\_sw\_cal/calib/cross\_cal/}]{2006ESASP.604..937S}. The photo-electric absorption was modeled as a combination of Galactic foreground absorption and an additional column density accounting for both the interstellar medium of the LMC and the intrinsic absorption by the source. The Galactic photo-electric absorption was set to a column density of N$_{\rm H{\rm , GAL}}$ = 6.44\hcm{20} \citep{1990ARA&A..28..215D} with abundances according to \citet{2000ApJ...542..914W}. The LMC-intrinsic column density $\rm N_{\rm{H},\rm{LMC}}$ was left as a free parameter with abundances of 0.49  for elements heavier than helium \citep{2002A&A...396...53R}. All the uncertainties were calculated based on a  $\Delta\chi^2$ statistic of 2.706, equivalent to a 90\% confidence level for one parameter of interest. 

The pulse-phase averaged spectra were first fitted with an absorbed power law, resulting in a reduced $\chi^2_{\rm{red}}=0.95$. We further tested models typically applied to the spectra of BeXRBs. In particular from the residuals of the fitted power-law model, there is evidence that adding a soft energy component ($\sim$1 keV) would improve the fit. The overall fit quality was marginally improved by the use of the {\tt mekal} (thermal component based on emission from optically thin plasma), {\tt bbody} or {\tt diskbb} (emission from an accretion disk consisting of multiple black-body components) models available in {\tt xspec}. The tested models with their best-fit parameters are summarised in Table \ref{tab:spectra}. However, all the thermal models improved the residuals at  energies around 1 keV. The combination of a power-law and a black-body component provided two different solutions with similar residuals. The first solution described a black body that mainly attributed for the softer X-rays, while the 
second solution (hereafter BBrad2) provided a black body of higher temperature (kT $\sim$ 2.8 keV) that contributed most of the flux at higher energies. 
The photon index of the power law for all the tested models was between 0.8 and 0.9 (apart from model BBrad2), which falls within the typical range of 0.6 to 1.4 as reported by \citet{2008A&A...489..327H}. 
The EPIC-pn data provide an indication of an emission line close to 6.4 keV. The line was fitted with a Gaussian profile (E=6.4$\pm$0.1 keV, $\sigma\sim0$) with equivalent width of $50\pm30$ eV.
The best-fit power-law model with the contribution of the low-temperature black body and the Fe line  is shown in Fig.\ref{fig:spec1}. 

Given the variations in the HR diagrams (see Fig. \ref{fig:pp}) we continued performing  phase-resolved spectroscopy using 5 phase bins. This was achieved with the use of the {\tt SAS} task {\tt phasecalc}, with zero phase defined as the date of the observation (MJD: 56314.0), the same as used in Fig.\ref{fig:pp}. For the phase-resolved analysis we used only the EPIC-pn data, because the frame time of the EPIC-MOS cameras is larger than the duration of each phase bin. The phased-resolved spectra were again fitted with a single power-law and  power-law plus thermal component models.

\begin{figure}
 \resizebox{\hsize}{!}{\includegraphics[angle=0,clip=]{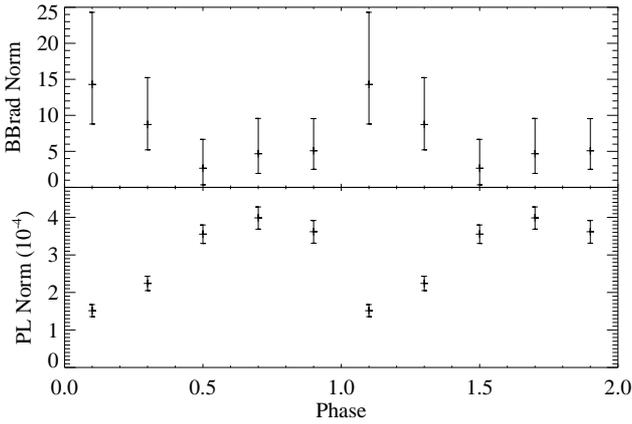}}
  \caption{Best fit values of the normalization parameters derived from phase-resolved spectroscopy. The absorption values were assumed to be constant. Error bars indicate the 90\% confidence level.}
  \label{fig:NORM2}
\end{figure}

It should be noted that the addition of the thermal component did not always improve the fit without producing unphysical parameters (e.g. black-body temperature of $>$100 keV). Given this behaviour we used a model with the same absorption for the five phases, linked the temperature of the thermal component and the power-law photon index, and let their normalizations vary. In Fig.\ref{fig:NORM2} we plot the normalization, therefore the strength, of each component versus the spin phase. 
The black-body component significantly contributes to the total flux during the first two phase bins (consistent with the worse fit quality found for the single power-law fit). The variation of the black-body intensity appears to be anti-correlated with that of the power-law component, although formally also a constant black-body flux can not be excluded.
When additionally allowing the absorption to vary with phase we note a similar behaviour of the normalizations. Again, there is some indication for a changing column density, but one more free fit parameter increases the errors on the individual parameters further, not allowing any firm conclusion about which parameters are principally variable.
Following the detection of an emission line near 6.4 keV in the phase averaged spectrum, we searched the phased spectra for similar features (see Table \ref{tab:spectra}).  
Because of the lower statistics in the phased spectra, we tested the significance of the line by fitting the model 
with a Gaussian line with zero width, stepping the line energy between 2 and 9 keV (100 steps), for all the phased spectra. The derived equivalent widths of the fitted lines were always less than 60 eV, except close to 6.4 keV. At centroid energies near 6.4 keV the equivalent width of the line exceeded 60 eV in two spectra (phase 0.2-0.4 and 0.6-0.8) suggesting a variation of the line intensity with pulse phase.

\begin{table*}
\caption[]{Spectral fit results.}

\begin{center}
\begin{tabular}{lccccccc}
\hline\hline\noalign{\smallskip}
\multicolumn{1}{l}{Model \tablefootmark{a}} &
\multicolumn{1}{c}{LMC \nh} &
\multicolumn{1}{c}{$\Gamma$} &
\multicolumn{1}{c}{Norm \tablefootmark{b}} &
\multicolumn{1}{c}{$\rm kT$} &
\multicolumn{1}{c}{Norm} &
\multicolumn{1}{c}{-} &
\multicolumn{1}{c}{$\chi^2_{\rm{red}}/{\rm dof}$ \tablefootmark{c}} \\
\multicolumn{1}{c}{} &
\multicolumn{1}{c}{[\oexpo{20}cm$^{-2}$]} &
\multicolumn{1}{c}{} &
\multicolumn{1}{c}{[\oexpo{-4}]} &
\multicolumn{1}{c}{[keV]} &
\multicolumn{1}{c}{} &
\multicolumn{1}{c}{-} &
\multicolumn{1}{c}{} \\

\noalign{\smallskip}\hline\noalign{\smallskip}
PL               & $3.4\pm1.0$  & $0.925\pm0.020$ & $3.52\pm0.09$& --& --& --&  0.953/975\\
\noalign{\smallskip}
PL+DiskBB        & $8.3^{+3}_{-2.4}$    & $0.83\pm0.04$   & $3.05^{+0.19}_{-0.21}$& $0.33^{+0.08}_{-0.07}$ & $1.1^{+2}_{-0.7}$ & --&  0.917/973  \\
\noalign{\smallskip}
PL+BBrad \tablefootmark{d}         & $5.3^{+2.4}_{-1.9}$    & $0.83\pm0.04$   & $3.05\pm0.18$& $0.24\pm0.04$ & $4.6^{+5}_{-2.1}$ & --&  0.914/973 \\
\noalign{\smallskip}
PL+BBrad2 \tablefootmark{d}  & $10.3^{+2.6}_{-2.4}$    & $1.35\pm0.13$   & $3.82^{+0.16}_{-0.15}$& $2.82^{+0.4}_{-0.26}$ & $0.007\pm0.003$ & --&  0.917/973 \\
\noalign{\smallskip}
PL+mekal        & $3.65^{+1.1}_{-1.0}$    & $0.86\pm0.03$   & $3.20^{+0.14}_{-0.13}$& $1.04\pm0.3$ & $9.9^{+5}_{-3}$\expo{-5} & --&  0.924/973 \\
\noalign{\smallskip}\hline
\noalign{\smallskip}\hline\noalign{\smallskip}

\multicolumn{1}{l}{PL+Gauss \tablefootmark{e}} &
\multicolumn{1}{c}{LMC \nh} &
\multicolumn{1}{c}{$\Gamma$} &
\multicolumn{1}{c}{Norm} &
\multicolumn{1}{c}{$\rm kT_{BB}$} &
\multicolumn{1}{c}{Norm} &
\multicolumn{1}{c}{Gauss \tablefootmark{f}} &
\multicolumn{1}{c}{$\chi^2_{\rm{red}}/{\rm dof}$}   \\
\multicolumn{1}{l}{Phase} &
\multicolumn{1}{c}{[\oexpo{20}cm]} &
\multicolumn{1}{c}{} &
\multicolumn{1}{c}{\oexpo{-4}} &
\multicolumn{1}{c}{keV} &
\multicolumn{1}{c}{} &
\multicolumn{1}{c}{keV/eV} &
\multicolumn{1}{c}{}  \\
\noalign{\smallskip}\hline\noalign{\smallskip}

0.0-0.2  & $3.55\pm0.02$ & $1.20\pm0.05$  & 2.86& --& --& $\sim$0 & 1.47/129  \\
0.2-0.4  & $\sim0$       & $0.88\pm0.04$  & 2.63& --& --& 6.38/227 & 1.14/160  \\
0.4-0.6  & $1.76\pm0.02$ & $0.75\pm0.04$  & 3.32& --& --& 6.38/50 & 1.11/202   \\
0.6-0.8  & $4.96\pm0.02$ & $0.89\pm0.04$  & 4.34& --& --& 6.38/97 & 1.07/210 \\
0.8-1.0  & $9.42\pm0.03$ & $1.02\pm0.05$  & 4.65& --& --& $\sim$0 & 1.16/181  \\
\noalign{\smallskip}\hline

\noalign{\smallskip}\hline\noalign{\smallskip}

\multicolumn{1}{l}{PL+BBrad \tablefootmark{g}} &
\multicolumn{1}{c}{LMC \nh} &
\multicolumn{1}{c}{$\Gamma$} &
\multicolumn{1}{c}{Norm} &
\multicolumn{1}{c}{$\rm kT_{BB}$} &
\multicolumn{1}{c}{Norm} &
\multicolumn{1}{c}{Radius \tablefootmark{h}} &
\multicolumn{1}{c}{$\chi^2_{\rm{red}}/{\rm dof}$}  \\
\multicolumn{1}{l}{Phase} &
\multicolumn{1}{c}{[\oexpo{20}cm]} &
\multicolumn{1}{c}{} &
\multicolumn{1}{c}{\oexpo{-4}} &
\multicolumn{1}{c}{keV} &
\multicolumn{1}{c}{} &
\multicolumn{1}{c}{km} &
\multicolumn{1}{c}{}  \\
\noalign{\smallskip}\hline\noalign{\smallskip}

0.0-0.2  & $4.5^{+2.4}_{-1.9}$       & 0.83$\pm0.05$   & 1.51$\pm0.16$ & 0.24$\pm0.03$ & $14.3^{+10}_{-6}$  & 18.9   & 1.19/534  \\
0.2-0.4  & $=$       & $=$   & 2.24$\pm0.19$ & $=$ & $8.7^{+7}_{-4}$    & 14.8   &  \\
0.4-0.6  & $=$       & $=$   & 3.55$\pm0.25$ & $=$ & $2.7^{+4}_{-2.3}$  & 8.2   &  \\
0.6-0.8  & $=$       & $=$   & 3.99$\pm0.3$  & $=$ & $4.7^{+5}_{-2.7}$  & 10.8  &  \\
0.8-1.0  & $=$       & $=$   & 3.62$\pm0.3$  & $=$ & $5.1^{+5}_{-2.6}$  & 11.3    &  \\
\noalign{\smallskip}\hline
0.0-0.2  &  $0.53^{+4.1}_{-0.53}$  & 0.80$\pm0.05$   & $1.41^{+0.18}_{-0.08}$& 0.276$^{+0.03}_{-0.04}$ &$7.5^{+7.5}_{ -2.2}$  & 13.7   & 1.19/530  \\
0.2-0.4  &  $1.0^{+4.2}_{-0.99}$  & $=$   & $2.13^{+0.21}_{-0.19}$& $=$ &$4.4^{+4.9}_{ -1.5}$  & 10.5   &  \\
0.4-0.6  &  $1.7^{+3.1}_{-1.7}$  & $=$   & $3.44^{+0.27}_{-0.27}$& $=$ &$0.87^{+2.3}_{-0.87}$  & 4.7   & \\
0.6-0.8  &  $4.3^{+3.3}_{-2.6}$  & $=$   & $3.8\pm0.3$           & $=$ &$3.5^{+3.9}_{ -2.0}$  & 9.3  & \\
0.8-1.0  &  $6.15^{+3.6}_{-2.9}$  & $=$  & $3.4\pm0.3$           & $=$ &$4.8^{+3.9}_{ -2.0}$  & 11.0    &  \\
\noalign{\smallskip}\hline

\multicolumn{8}{l}{
  \tablefoot{
  \tablefoottext{a}{For definition of spectral models see text.}
  \tablefoottext{b}{Photons keV$^{-1}$cm$^{-2}$s$^{-1}$ at 1 keV.}
  \tablefoottext{c}{Degrees of freedom (dof).}
  \tablefoottext{d}{The black-body model could both be fitted with a low or a high temperature component, see text for more details.}
  \tablefoottext{e}{Parameters for phase-resolved analyses, based on a power law plus Gaussian line.}
  \tablefoottext{f}{Center and equivalent width of Fe line if present.}
  \tablefoottext{g}{Parameters for phase resolved analyses, based on a power law plus a thermal component. }
  \tablefoottext{h}{Radius of the emitting area, derived from the black-body normalization.}
  }}

\end{tabular}
\end{center}
\label{tab:spectra}
\end{table*}

\subsection{X-ray long-term variability}
\label{sec-Xvar}
From the spectral analysis of the \xmm ToO observation of \rxp, we calculated a flux of $6.00^{+0.2}_{-0.16}$\ergcm{-12} in the 0.3--10.0 keV band. This translates into an absorption corrected luminosity of 1.79\ergs{36} for a distance to the LMC of 49 kpc \citep{2013ApJ...764...84I}.
During the \xmm observation performed on 2004 January 17, the source was located over a gap of the EPIC-pn camera and near bad columns of the EPIC-MOS cameras and was marginally detected at a flux level of 2.15\ergcm{-14}. 
Based on the \xmm observations we estimate a long-term X-ray variability factor of at least 280 for \rxp. Both the lowest and highest fluxes have been measured from \xmm observations, while the \swift flux was always between these two values. The ROSAT measured flux was near the \xmm ToO value, however it does not provide any additional constraints, given that the two observatories cover different energy bands and  that the photo-electric absorption will have a large effect on the soft ROSAT energy band which may introduce large systematic uncertainties. 


\section{Analysis and results of optical data}
\label{sec:datareduction_optical}

\subsection{Optical counterpart}
The most accurate X-ray position of \rxp was derived from \xmm ToO data (Sect. \ref{sec-xobs}). Within the 3$\sigma$ error circle we found a V=14.12 mag star as only likely counterpart at a distance of $\sim$0.8\arcsec\ to the X-ray position. This star is included in the 2MASS catalogue (2MASS\,J0520299-6931559). This confirms the previously suggested counterpart by \citet{1994PASP..106..843S}, as the real companion of \rxp.

\begin{figure}
 \resizebox{0.97\hsize}{!}{\includegraphics[angle=0,clip=]{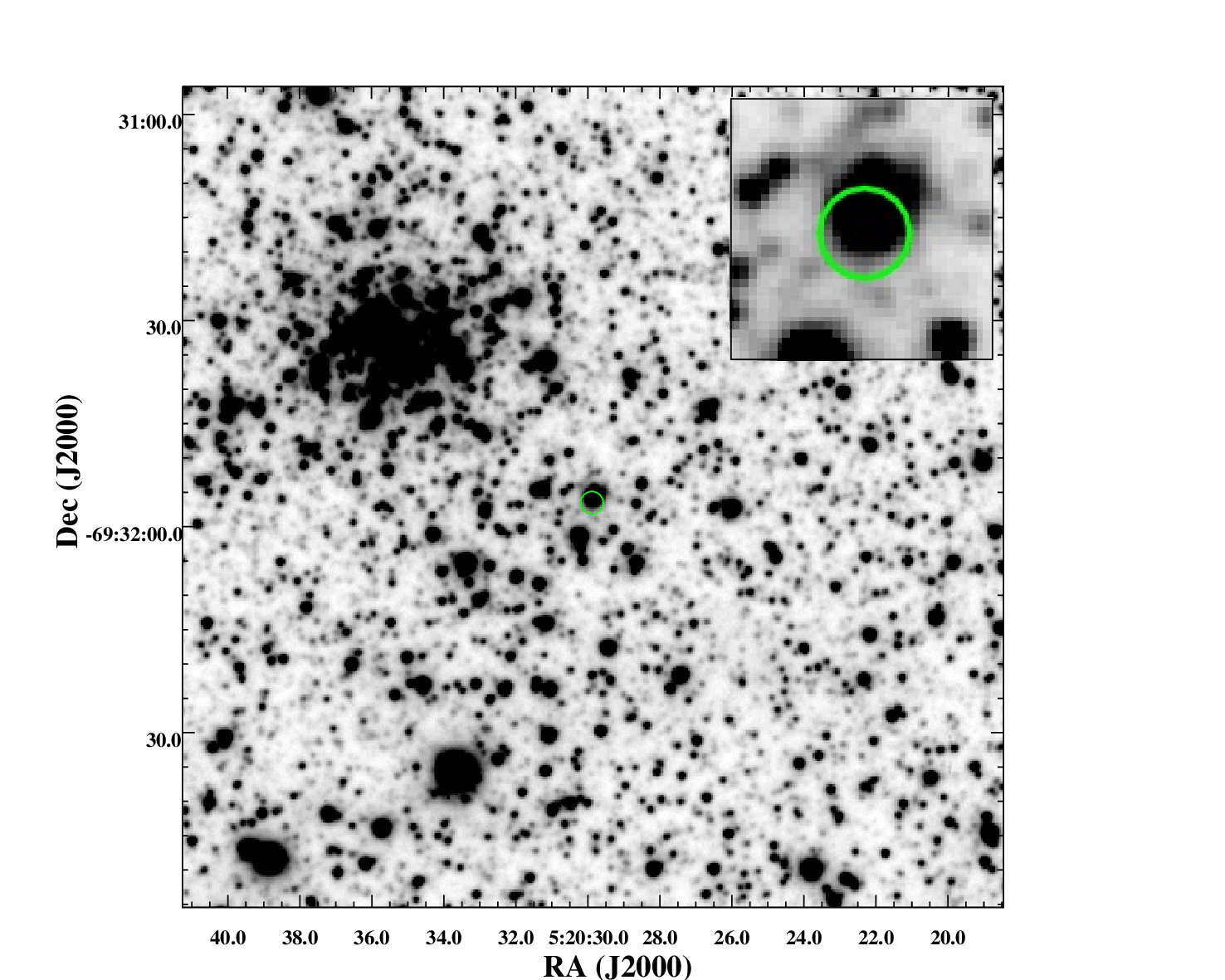}}
  \caption{Finding chart of \rxp. The image is a product of the OGLE project. The 3$\sigma$ X-ray positional uncertainty is marked with a green circle. On the upper left part of the image, we find the star cluster NGC\,1926, at a distance of $\sim$0.64\arcmin.}
  \label{fig:fchart}
\end{figure}

\begin{figure}
 \resizebox{\hsize}{!}{\includegraphics[angle=0,clip=]{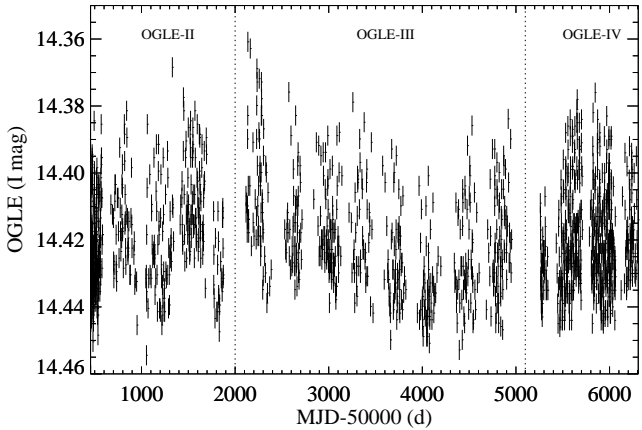}}
  \caption{OGLE I band light curve of \rxp. Data from different OGLE phases are separated with vertical lines.}
  \label{fig:ogle_lc_all}
\end{figure}

\subsection{OGLE light curve}
\label{sec-ogle}

The Optical Gravitational Lensing Experiment (OGLE) started its initial observations in 1992 \citep{1992AcA....42..253U} and continues observing till today, with its most recent data release (OGLE III) in 2008 \citep{2008AcA....58...69U}. Observations were made with the 1.3 meter Warsaw telescope at Las Campanas Observatory. Images are taken in the V and I filter pass-bands, while the data reduction is described in \citet{2008AcA....58...69U}. 

In the present work, we used the OGLE I and V band magnitudes for the counterpart of \rxp that were taken before MJD=56600 (see Table \ref{tab:ogle}). The latest I band light curve is seen in Fig. \ref{fig:ogle_lc_all}, where the different OGLE phases have been normalized to the same mean value to account for calibration offsets. The data can be downloaded from the OGLE-IV real time monitoring of X-ray sources web-page\footnote{{\tt XROM}: http://ogle.astrouw.edu.pl/ogle4/xrom/xrom.html}. 

\begin{figure}
\resizebox{\hsize}{!}{\includegraphics[angle=0,clip=]{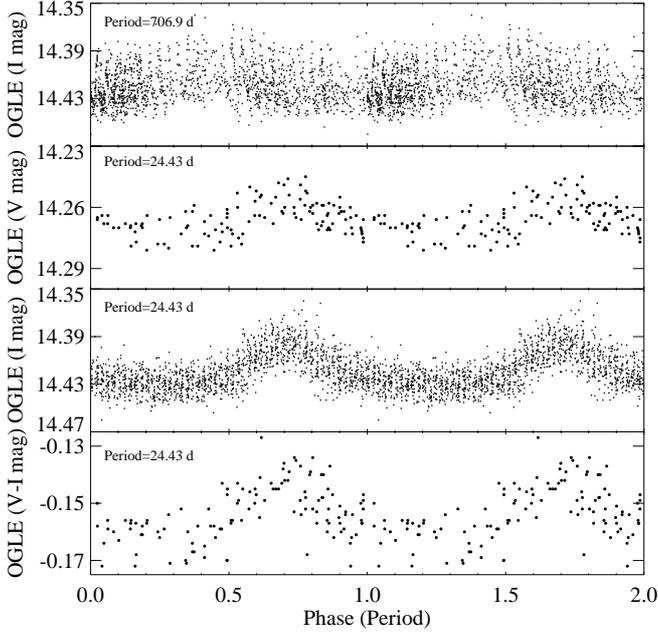}}
  \caption{OGLE I band, V band and V-I light-curve profiles, folded for a period of 24.43 d and 706.9 d. The I band curve consists of data from phases II, III and IV, while the V band we only had data from phase IV. All the data are folded using as a zero-phase the first available OGLE data point (MJD: $\sim$50455.67).}
  \label{fig:fold_ogle}
\end{figure}

\begin{figure}
 \resizebox{\hsize}{!}{\includegraphics[angle=0,clip=]{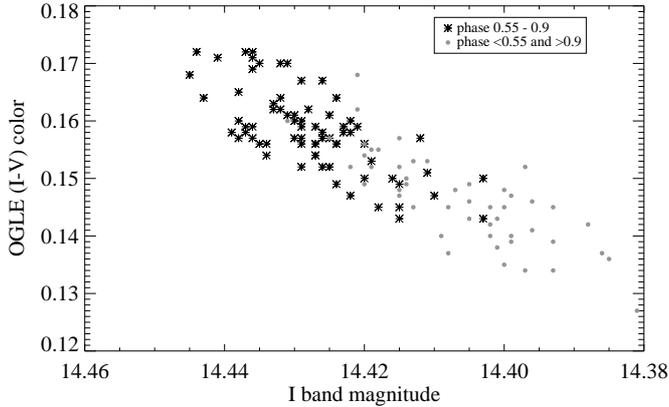}}
  \caption{Colour magnitude diagram of \rxp. The OGLE data are divided into two groups according to the phase derived from the 24.43 d period.}
  \label{fig:color_ogle}
\end{figure}

From the different OGLE data sets of the I band magnitudes we computed periodograms by using the Lomb-Scargle (LS) algorithm \citep{1982ApJ...263..835S,1986ApJ...302..757H}. We searched for periodicities up to half of the total OGLE observing period, which at the time of the analysis was $\sim$5851 d.
Since the LS algorithm does not determine the period error, we use the bootstrap method to estimate the period uncertainties. A random sampling is drawn from the light curve (one epoch can be drawn multiple times) and the period for that sampling is determined with the LS algorithm. This procedure is performed 1000 times for each light curve (OGLE phase), so that the 1 $\sigma$ uncertainty can be determined from the resulting distribution.
Based on the combined OGLE data, we derived a new improved period of $24.4302\pm0.0026~{\rm d}$ for the optical counterpart, while the average luminosity level was nearly constant for the last 15 years
(variability less than 0.1 mag). 
By analysing the available V band data we found a similar period (24.22 d).
We also analysed the I band data from the three different OGLE phases separately, to search for possible differences between them.
This resulted in the detection of slightly different periods for the three OGLE phases. We also detected short periodicities with values 1.04 d, 0.958 d, 0.509 d and 0.49 d which can be associated to the 0.5 and 1 day aliases of the 24.43 d period ($1/(0.5^{-1}\pm24.43^{-1})$ and $1/(1^{-1}\pm24.43^{-1})$) and are not associated with the NRP phenomenon. 
When analysing the complete I band light curve from the three available OGLE phases we found an indication of a long period of $\sim$706.9 d. Such super-orbital periods have been reported from other BeXRB systems in the MCs and they might be related to the formation and depletion of the circumstellar disk around the Be star \citep{2011MNRAS.413.1600R}. 
The results of our temporal analysis are listed in Table \ref{tab:ogle}.
For the period where both I and V band measurements are available we are able to follow the evolution of the colour of the counterpart of \rxp. To calculate the I-V colour we used only the measurments less than one day apart. 
In Fig. \ref{fig:fold_ogle} we present the folded light curve for the OGLE I and V band magnitudes and the V-I colour of the optical counterpart of \rxp.
In Fig. \ref{fig:color_ogle} we present the colour magnitude diagram for the period between 55260 and 56600 (MJD).

\begin{table}[h]
\caption{OGLE data for the optical counterpart of \rxp.}
\begin{center}
\begin{tabular}{lccc}
\noalign{\smallskip}\hline\noalign{\smallskip}
OGLE\tablefootmark{a} & Data points & Period (d) & Power\tablefootmark{b}\\
\noalign{\smallskip}\hline\noalign{\smallskip}
\multicolumn{4}{l}{I band}\\
II & 510 & 24.440$\pm$0.011 & 117 \\
III & 489 & 24.421$\pm$0.006 & 97 \\
IV & 590 & 24.367$\pm$0.018  & 149 \\
Total & 1589 & 24.4302$\pm$0.0026 & 350 \\
& & 706.9$\pm3$ & 60 \\
\multicolumn{4}{l}{V band}\\
IV & 127 & 24.22$\pm0.04$ & 26.6 \\
\noalign{\smallskip}\hline\noalign{\smallskip}

\multicolumn{4}{l}{
  \tablefoot{
  \tablefoottext{a}{The OGLE catalogue entries are lmc\,sc6.66519, lmc\,100.1.19283, lmc\,503.11.88150 for phases II, III and IV.}
  \tablefoottext{b}{A power of 9.8 for the I band and 13 for the V band corresponds to a confidence level of 99\%.}
  }}
\end{tabular}
\end{center}
\label{tab:ogle}
\end{table}

\section{Discussion}
 \label{discussion}
 
 \begin{figure}
 \resizebox{\hsize}{!}{\includegraphics[angle=0,clip=]{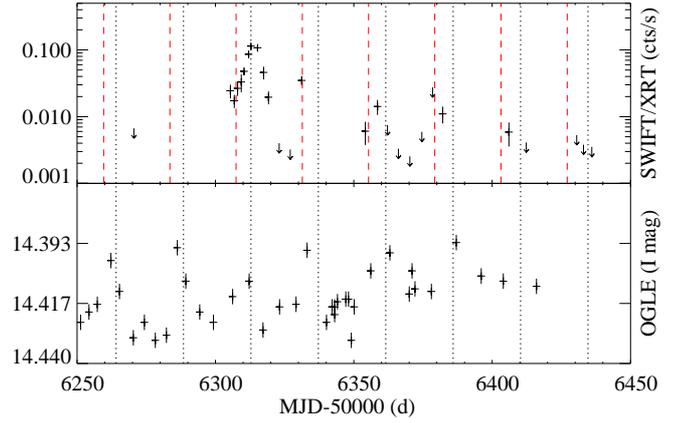}}
  \caption{Optical and X-ray light curve of \rxp. Top panel: \swift/XRT count rates (0.3--10 keV band), 2$\sigma$ upper limits are plotted with arrows. The vertical black dotted lines represent the 24.43 d optical period phased on the detected X-ray maximum, while the red dashed lines are spaced based on the 23.93 d orbital solution of \citet{2014ATel.5856....1K} (phased at the mean $T_{\rm 90}$ time). Bottom panel: OGLE-IV I band optical light curve.}
  \label{fig:swiftlc}
\end{figure}

Analysing the EPIC data of our \xmm ToO observation of \rxp, performed on 2013 January 22 during the maximum of a type I outburst, we detected coherent X-ray pulsations with a period of 8.035331(15) s (1$\sigma$). During the outburst, the X-ray spectrum is fitted best by an absorbed power law ($\Gamma=0.83$) plus a low-temperature black-body (kT $\sim0.24$ keV) model. From phase-resolved spectroscopy, we found that the spectral shape changes with spin phase. This modulation can be successfully modelled by the same model used for the phase-averaged spectrum, but with normalizations for the different components that vary with phase.   

It has been suggested that the presence of a soft spectral component is a common feature intrinsic to X-ray pulsars, which is related to the total luminosity of the source \citep{2004ApJ...614..881H}. For sources with higher luminosity ($L_X>$\oergs{38}) it could be a result of reprocessed hard X-rays from the neutron star by optically thick accreting material. While in less luminous sources ($L_X<$\oergs{36}), a soft excess could be due to emission by photoionised or collisionally heated gas or thermal emission from the surface of the neutron star. Either or both of these types of emission can be present for sources with intermediate luminosity.

In the last decade, a thermal soft excess has been detected in a number of BeXRB pulsars. In several persistent systems, with luminosities between \oergs{34} and \oergs{36}, the X-ray spectra can be fitted by a power law and a black body with temperature values between 1.1 and 1.8 keV \citep[e.g.][]{2013arXiv1301.5120L,2013MNRAS.436.2054B}. Some other systems, commonly observed during outburst, also show evidence of thermal emission but with much lower temperatures around 0.3 keV \citep[e.g.][]{2008A&A...484..451H,2011A&A...527A.131S}.
In the case of \rxp, we note that the system was observed during an outburst, therefore we tried to test thermal models with a larger range in temperature.
We find that although, a black-body component of $\sim$2.8 keV yields an acceptable fit quality (see PL+BBrad2 in Table \ref{tab:spectra}), it also produces some questionable properties for the system. In the case of the higher temperature black-body model, the power-law component accounts for the softer X-rays and the black-body component contributes the majority of the flux at high energies, which is difficult to explain by heating or reprocessing effects. 

The change in the normalizations of the power-law and black-body components with spin phase might be explained by a simple picture. First, the maximum radius ($\sim$20 km) of the black-body emission region argues against a hot spot on the neutron stars surface. More likely it could be a region in the inner part of an accretion disk. If we imagine that the accretion column originating from the disk and terminating on the NS surface, depending of the geometry of the system with respect to the observer, it might be possible for the accretion column to partially obscure the NS and/or the inner part of the accretion disk. However, in order to deeper study this behaviour observations with better statistics are required. 

A total X-ray long-term variability factor of $\sim$280 was derived from the available \xmm observations of \rxp.
In Fig. \ref{fig:swiftlc}, we present the X-ray light curve from our \swift monitoring together with the optical I-band light curve from the same period. 
The \swift/XRT fluxes (detections and upper limits) are all within the range found from the \xmm observations. 
There is evidence for three additional outbursts after the first one, these events seem to follow the periodic behaviour of the optical counterpart and the $\sim$24.4 d period. Unfortunately, the available X-ray coverage does not allow to make this case stronger.

In Fig. \ref{fig:fold_ogle}, we show that a periodic behaviour is seen in both the I and V band light curve as well as in the V-I colour. The colour periodicity is also translated as a larger variability in the I than in the V band magnitude. In Fig. \ref{fig:color_ogle} we show the evolution of the colour of the optical counterpart of \rxp during an approximately three year period. From the relative positions of the data points from different phases, it can be seen that the colour of the system correlates well with its luminosity, being redder when brighter. In similar studies concerning the optical properties of Be stars, a colour magnitude variability pattern that follows an anticlockwise path in the colour magnitude diagram (same axes as Fig. \ref{fig:color_ogle}) has been reported \citep[e.g.][]{2006A&A...456.1027D,2012MNRAS.424..282C}. \citet{2011MNRAS.413.1600R} studied the colour magnitude diagrams of 31 BeXRB systems finding that only 4 or 5 of them become bluer when brighter. They proposed that 
this 
is related with the orientation by which we observe the system, concluding that in systems with bluer-brighter relation the decretion disk obscures the Be star, while the redder-brighter relation describes a system where the disk is observed face-on. 

In almost all these studies the systems showed a much higher variability in their magnitudes and exhibit some kind of sharp outburst rather than the case of a flat light curve of \rxp. 
In systems like XMMUJ010743.1-715953 \citep{2012MNRAS.424..282C} the colour magnitude evolution pattern presumably reflects the growth of the disk as a large cooler component of the Be star. In the case of \rxp we clearly see that this colour-magnitude relation does not reflect the complete loss and build up of the disk, since we only see a periodic variability of the magnitude with small amplitude.
This behaviour indicates that a similar, although less intense, disk evolution happens within the orbital period of the system and is most likely related to the NS motion. 
The NS orbiting the Be star is perturbing the decretion disk, likely via tidal truncation \citep{2001A&A...377..161O}, resulting in an increased emission from the redder component of the system (e.g. cooler disk).

\citet{2014ATel.5856....1K} have recently analysed the pulse-period evolution provided by the GBM (Fermi Gamma-ray Burst Monitor) pulsar project to derive the orbital parameters of the binary system by modeling the Doppler shift due to the orbital motion. 
Including a luminosity-dependent intrinsic spin-period derivative, they found an orbital solution with a period of $23.93\pm0.07$ d, a projected semi-major axis of $a\sin{i}=107.6$ ls, an eccentricity $e=0.0286$ and the time at which the delay in the arrival of the pulse is maximized $\rm T_{\rm 90}=2456666.91 (JD)$. This period is significantly shorter than the period derived from the OGLE photometry, but should be considered as the most accurate solution for the orbital motion of the binary.  
By using Kepler's 3$^{\rm rd}$ law (neglecting the small eccentricity) and typical values for the masses of the system \citep[1.4 and 23 $M_{\odot}$,][]{1996ApJ...460..914V}, we estimate the actual orbit of the system to be $\sim$2.3 times larger than the projected semi-major axis. That suggests that the orbital plane of the NS has an inclination of about $26^{\circ}$ (where $0\,^{\circ}$ indicates a face-on system).
   
It is not uncommon for BeXRB systems to exhibit small differences in the periods derived from the optical and from pulse timing. \citet{2011MNRAS.416.1556T} reported that for  two SMC pulsars, SXP2.37 and SXP6.85. Their orbital periods derived from pulse timing are smaller than the optical (by 0.22 d and 3.9 d respectively). A likely explanation for this difference is that the variability in the optical data is a combined product of two motions, the Keplerian orbit of the NS around its massive companion and the rotation of the circumstellar disk. In Table \ref{tab:ogle} we see that the periodicity values derived from OGLE phases II and IV are statistically different (3.5 $\sigma$). This might be a result of small changes in the rotation speed or matter distribution of the decretion disk.
The observed outbursts in the optical  might be a combined result of the co-rotation of the Be disk in the same direction with the NS orbit, and inhomogeneities in the disk. Inhomogeneities are known from cyclic variations observed in emission line profiles of other systems. \citet{1991PASJ...43...75O}, explained these variations as a result of one armed oscillations in the equatorial disks of the Be stars. \citet{2011MNRAS.413.1600R} confirmed that the long-term variation in the optical light curves of BeXRB systems is related to the behaviour of 
the Be circumstellar disk.  From the OGLE light curve of the optical counterpart of \rxp, there is evidence for a super-orbital period. Both the amplitudes of this long-term periodicity and the total variability of the source, are even smaller than the variability within the orbital period. This suggests a Be disk observed nearly face on, consistent with what we derived above.

Knowing that the BeXRB population correlates with star-formation regions with young ages it is likely that the system originated from a nearby cluster.
\rxp is located near the star cluster \object{NGC\,1926} at a projected angular distance of 0.65\arcmin~(see Fig. \ref{fig:fchart}). NGC\,1926 has an age of $31^{+30}_{-14}$ Myr \citep{2012ApJ...751..122P}. 
For the case of BeXRBs in the SMC a convincing link between the spatial position of the binary systems and near young star clusters has been reported by \citet{2005MNRAS.358.1379C}. Their analysis leads to an average space velocity of $\sim$30 \kms\ for the binary systems which is proposed to arise from a supernova kick.
Assuming a value of 5 Myr for the most likely maximum lifetime of the Be star after the NS has been formed \citep{1977ApJ...214L..19S} we can estimate a lower limit for the runaway velocity imposed to the system by the  supernova kick. In our case this leads to a velocity of 1.8 \kms. Knowing that the systems true motion has a random direction to our line of sight the true velocity could be larger.  On a statistical basis this value is smaller than the average velocity reported for BeXRB systems in the SMC, or in our galaxy \citep[e.g. $19\pm8$ \kms,][]{2000A&A...364..563V}. An independent argument in favour of the small kick velocity is the small eccentricity of the binary system.

\rxp adds to the important sample of BeXRBs with known spin period, orbital period and eccentricity.
The spin and orbital periods are consistent with the Corbet relation \citep{1984A&A...141...91C} and the low eccentricity of the system is expected for systems with small spin periods \citep{2011Natur.479..372K}. According to their interpretation, BeXRB could originate from two different types of supernova progenitors, with electron-capture supernovae preferentially producing systems with short spin period, short orbital periods and low eccentricity like \rxp.       

We finally note that in January 2014 \rxp started a major outburst reaching a luminosity of 1.9\ergs{38} \citep{2014ATel.5760....1V}. This is almost 100 times brighter than the outburst of January 2013. The outburst still continues at the time of writing and will be the subject of a subsequent paper.

\section{Conclusion}
\label{conclusion}   
The analysis of our \xmm ToO observation of \rxp revealed X-ray pulsations with a period of $8.035331(15)$~s. The X-ray spectrum is best fitted by an absorbed power-law with a spectral index of $\sim$0.83 plus a low-temperature black-body component with temperature of $\sim$0.24 keV. The detailed pulse-phase resolved analysis of its spectral properties reveals a significant change in its spectral shape. Assuming the same spectral model the changes with pulse phase can be described by variations in the black-body and power-law intensities. A possible anti-correlation of black-body and power-law flux (possibly due to geometrical effects) requires confirmation with data of higher statistical quality. Our analysis of the OGLE light curve of the optical counterpart confirmed an optical period of 24.43 d. Our results confirm \rxp as a BeXRB with a NS primary, making it the 15$^{th}$ known HMXB pulsar in the LMC.

\begin{acknowledgements}
We would like to thank the referee Prof. M.\,J.\, Coe for his suggestions that helped improve the manuscript.  
The \xmm\ project is supported by the Bundesministerium f\"ur Wirtschaft und
Technologie\,/\,Deutsches Zentrum f\"ur Luft- und Raumfahrt (BMWi/DLR, FKZ 50 OX
0001) and the Max-Planck Society. The OGLE project has received funding from the
European Research Council under the European Community's Seventh Framework
Programme (FP7/2007-2013)\,/\,ERC grant agreement no.\,246678 to A.\,U.
We thank the SWIFT team for accepting and carefully scheduling the target of opportunity observations, and we acknowledge the use of public data from the SWIFT data archive. 
G.\,V., P.\,M. and R.\,S acknowledge support from the BMWi/DLR grants FKZ 50 OR 1208, 1201 and 0907 respectively.
\end{acknowledgements}

\bibliography{RXJ0520}

\onecolumn
\begin{appendix}
\label{appenA}
 \section{X-ray observation logs}

 \begin{center}
 \begin{longtable}{lcccccc}
\caption{X-ray observations of \rxp.}\\
     \hline\hline\noalign{\smallskip}
     \multicolumn{1}{c}{ObsID} &
     \multicolumn{1}{c}{MJD} &
    \multicolumn{1}{c}{Instrument\tablefootmark{a}} &
    \multicolumn{1}{c}{Off-axis angle\tablefootmark{b}} &
     \multicolumn{1}{l}{Net Exp} &  
     \multicolumn{1}{r}{Net Count rates\tablefootmark{c,d}} &
     \multicolumn{1}{c}{$F_{\rm x}$\tablefootmark{e}}  \\

     \multicolumn{1}{l}{} &
     \multicolumn{1}{c}{[$T_{\rm start}$]} &
     \multicolumn{1}{l}{} &
     \multicolumn{1}{c}{[\arcmin]} &
     \multicolumn{1}{c}{[ks]} &   
     \multicolumn{1}{c}{[$10^{-2}$ cts s$^{-1}$]} &
     \multicolumn{1}{c}{[erg s$^{-1}$ cm$^{-2}$]}  \\
     \noalign{\smallskip}\hline\noalign{\smallskip}
\endfirsthead
\caption{Continued.}\\

     \hline\hline\noalign{\smallskip}
     \multicolumn{1}{c}{ObsID} &
     \multicolumn{1}{c}{MJD} &
     \multicolumn{1}{c}{Instrument} &
      \multicolumn{1}{c}{Off-axis angle} &
     \multicolumn{1}{l}{Net Exp} &  
     \multicolumn{1}{r}{Net Count rates} &
     \multicolumn{1}{c}{$F_{\rm x}$}  \\

     \multicolumn{1}{l}{} &
     \multicolumn{1}{c}{[$T_{\rm start}$]} &
     \multicolumn{1}{l}{} &
     \multicolumn{1}{c}{[\arcmin]} &
     \multicolumn{1}{c}{[ks]} &   
     \multicolumn{1}{c}{[$10^{-2}$ cts s$^{-1}$]} &
     \multicolumn{1}{c}{[erg cm$^{-2}$ s$^{-1}$]}  \\
     \noalign{\smallskip}\hline\noalign{\smallskip}
\endhead

\hline 
\multicolumn{7}{l}{
  \tablefoot{
  \tablefoottext{a}{Observation setup for XMM: full-frame mode (ff) and photon-counting mode (pc). For \xmm, the medium filter (m) was used.}
  \tablefoottext{b}{Off-axis angle under which the source was observed.}
  \tablefoottext{c}{Net count rate for the (0.2 -- 10.0) keV band of \xmm.}
   \tablefoottext{c}{\swift XRB count rates in the (0.2 -- 10.0) keV band corrected for Vignetting, as estimated with the SOSTA XIMAGE command.}
  \tablefoottext{e}{X-ray flux in the (0.3 -- 10.0) keV band, derived from the best fit spectral model.}
  \tablefoottext{f}{Source intrinsic X-ray luminosity in the (0.3 -- 10.0) keV band (corrected for absorption)
for a distance to the LMC of 50 kpc \citep{2005MNRAS.357..304H}.}
  }
}
\endfoot
\endlastfoot

         XMM         &  &  &    &  &     \\
        0690750901   & 56266.1 &  EPIC-pn  & 10.9     &  22.24 & 0.94$\pm$0.15 &   7$\times10^{-14}$  \\
                     &     &  EPIC-MOS1   & 10.0   &  26.37 & 0.18$\pm$0.0006 & -  \\
                     &     &  EPIC-MOS2    & 10.5  &  26.39 & 0.14$\pm$0.0007 & -  \\

        0701990101  &  56314.0   &  EPIC-pn & 1.1    &  16.82 & 101.4$\pm$0.8 &   6.0$\times10^{-12}$  \\
                    &     &  EPIC-MOS1       & 0.17   &  20.3  & 33.6$\pm$0.5 & -  \\
                    &     &  EPIC-MOS2      & 1.2    &  20.31 & 33.3$\pm$0.4 & -  \\                 
      
     \noalign{\smallskip}\hline\noalign{\smallskip}
        Swift         &  &  &  &  &     \\                            
00045467001 & 56270.5 & &  11.6  &  1.12  & $<$1.0          & $<$7.7$\times10^{-13}$  \\ 
00045478001 & 56305.4 & &  8.0   &  1.65  & 2.4$\pm$0.6   & 18.4$\times10^{-13}$    \\
00032671001 & 56306.8 & &  2.3   &  1.97  & 1.7$\pm$0.4   & 13.1$\times10^{-13}$     \\
00032671002 & 56308.0 & &  1.9   &  1.28  & 2.7$\pm$0.6   & 20.7$\times10^{-13}$     \\
00045479001 & 56309.3 & &  7.2   &  0.70  & 3.3$\pm$1.0   & 25.4$\times10^{-13}$    \\
00032671003 & 56310.3 & &  2.0   &  2.32  & 4.8$\pm$0.6   & 36.9$\times10^{-13}$     \\
00045479002 & 56312.0 & &  8.6   &  2.91  & 8.6$\pm$0.8   & 66.1$\times10^{-13}$    \\
00045478002 & 56312.8 & &  8.5   &  1.99  & 11.4$\pm$1.1  & 87.6$\times10^{-13}$     \\
00032671004 & 56315.2 & &  3.7   &  2.12  & 10.8$\pm$1.0  & 83.0$\times10^{-13}$     \\
00045478003 & 56317.4 & &  8.7   &  1.03  & 4.6$\pm$1.0   & 35.4$\times10^{-13}$    \\
00032671005 & 56319.2 & &  2.3   &  1.84  & 2.0$\pm$0.4   & 15.4$\times10^{-13}$     \\
00032671006 & 56323.0 & &  2.8   &  2.03  & $<$0.4          & $<$3.1$\times10^{-13}$      \\
00032671007 & 56327.0 & &  2.8   &  2.15  & $<$0.32         & $<$2.5$\times10^{-13}$     \\
00032671008 & 56331.2 & &  1.9   &  1.92  & 3.5$\pm$0.6   & 26.9$\times10^{-13}$    \\
00032671009 & 56354.2 & &  0.07  &  2.21  & 0.61$\pm$0.23 & 4.7$\times10^{-13}$      \\
00032671010 & 56358.6 & &  2.2   &  1.92  & 1.4$\pm$0.4   & 10.8$\times10^{-13}$    \\
00032671011 & 56362.2 & &  0.6   &  2.02  & $<$0.64         & $<$4.9$\times10^{-13}$     \\
00032671012 & 56366.1 & &  2.9   &  1.82  & $<$0.32         & $<$2.5$\times10^{-13}$     \\
00032671013 & 56370.2 & &  2.1   &  2.30  & $<$0.26         & $<$2.0$\times10^{-13}$     \\
00032671014 & 56374.6 & &  1.7   &  1.85  & $<$0.6          & $<$4.6$\times10^{-13}$     \\
00032671015 & 56378.5 & &  2.9   &  0.249 & $<$0.28         & $<$2.2$\times10^{-13}$     \\
00032671016 & 56382.1 & &  0.6   &  2.17  & 1.1$\pm$0.3   & 8.5$\times10^{-13}$      \\
00032671017 & 56406.0 & &  1.3   &  2.00  & 5.87$\pm$0.23 & 45.1$\times10^{-13}$     \\
00032671019 & 56412.3 & &  2.0   &  2.12  & $<$0.4          & $<$3.1$\times10^{-13}$     \\
00032671020 & 56430.5 & &  0.7   &  1.00  & $<$0.53         & $<$4.1$\times10^{-13}$     \\
00032671021 & 56433.0 & &  3.0   &  1.85  & $<$0.38         & $<$2.9$\times10^{-13}$     \\
00032671022 & 56436.0 & &  2.2   &  2.16  & $<$0.35         & $<$2.7$\times10^{-13}$     \\                                

      \noalign{\smallskip}\hline\noalign{\smallskip} 

\multicolumn{7}{l}{
  \tablefoot{
  \tablefoottext{a}{Observation setup for \xmm: EPIC-pn in full-frame mode with thin filter, EPIC-MOS in full-frame mode with medium filter. All \swift observations are made with XRT instrument in photon-counting mode. }
  \tablefoottext{b}{The off-axis angle of individual detections given (\arcmin).}
  \tablefoottext{c}{Net count rate for the 0.3 -- 10.0 keV band of \xmm.}
   \tablefoottext{d}{\swift/XRT count rates in the 0.3 -- 10.0 keV band corrected for telescope vignetting. For the non-detections we list the 2$\sigma$ upper limits.}
  \tablefoottext{e}{X-ray observed flux in the 0.3 -- 10.0 keV band, derived from the best fit spectral model. The \swift fluxes were derived from the XRT count rates assuming the best-fit absorbed power-law model derived from the \xmm ToO spectra. }
  }
}
 \label{tab:xray-obs}

 \end{longtable}
\end{center}

\end{appendix}

\end{document}